\documentclass[aps,pre,twocolumn]{revtex4}
\usepackage{ae}
\usepackage{graphicx}

\begin{document}  
\narrowtext

\noindent
{\bf Comment on: Thermostatistics of Overdamped Motion of Interacting Particles}

In a recent Letter~\cite{An10} Andrade {\it et al.} argued that overdamped particles in contact with a reservoir at 
$T=0$ ``obey Tsallis statistics with
entropic index $\nu=2$''.  To justify this, Andrade {\it et al.} performed a simulation of classical 
particles in  two dimensions, interacting through a pair potential 
$V(r)=q G({\bf x_1},{\bf x_2})$, where $G({\bf x_1},{\bf x_2})=q K_0( |{\bf x_1}-{\bf x_2}|/\lambda)$, and 
$K_0$ is a modified Bessel function
of order zero.  This potential models  
a vortex-vortex interaction in a type II superconductor, 
where $q$ is the vortex strength and $\lambda$ is the effective London penetration length. 
The particles were confined in a 1d parabolic trap with a potential  $W(x)= \frac{\alpha}{2} x^2$, while a periodic
boundary condition was used in the $y$ direction (periodicity $L_y$).  
The simulation was performed using  overdamped molecular dynamics (MD)
with the thermostat set at $T=0$.  The resulting stationary density profile was then {\it fitted} to a distribution that
maximizes the Tsallis entropy of index  $\nu=2$.

The idea that non-extensive Tsallis statistics is relevant for particles at $T=0$
is very surprising and, {\it if correct},  would certainly justify publication in Physical Review Letters.
However, prior to discarding the standard statistical mechanics one 
should see what it has to say on this matter. 

Within traditional thermodynamics,
if a classical system is placed in contact with a temperature reservoir at $T=0$ it will loose all its kinetic energy and
collapse to the ground state  --- the minimum of the potential energy.  
In the ground state the net force on each particle vanishes.  We will now show that this is precisely
what happens for the system studied by Andrade {\it et al.}  

From now on we will measure all lengths in units of $\lambda$. We will work in 
thermodynamic limit $N \rightarrow \infty$, at fixed total vertex strength, $q^2 N=1$. 
We note first that the potential produced by a particle
located at $\bf x_1$ satisfies 
\begin{equation}
 \nabla^2 G({\bf x},{\bf x_1}) - G({\bf x},{\bf x_1}) =- 2 \pi q \delta({\bf x} - {\bf x_1}) \,.  
\label{pois}
\end{equation}
Taking into account the
periodicity in the $y$-direction, this equation can be solved exactly to yield
\begin{equation}
G({\bf x}; {\bf x_1})= \frac{\pi q}{L_y} \sum_{m=-\infty}^{\infty} e^{\frac{2 \pi m i}{L_y}(y-y_1)} \frac{e^{-\gamma_m |x-x_1|}}{\gamma_m} \, ,
\label{green}
\end{equation}
where $\gamma_m=\sqrt{1+ 4 \pi^2 m^2/L_y^2}$. Furthermore, symmetry requires that at equilibrium
the total potential inside the system is 
a function of $x$-coordinate only, $\varphi(x)$.  The force balance on each particle 
then reduces to $q \varphi'(x) =-\alpha x$.  In view of eq. (\ref{pois}),  $\varphi(x)$
must also satisfy the inhomogeneous Helmholtz equation, $\varphi''(x) - \varphi(x) = -2 \pi q \rho(x)$, 
from which we conclude that the particle density $\rho(x)$ is a restricted parabolic function,
%
\begin{equation}
\rho(x)=\frac{\alpha}{4 \pi q^2} \left(x_m^2-x^2 \right) \Theta(x_0^2 - x^2) \,,
\label{den}
\end{equation}
where $\Theta$ is the Heaviside step function. 
Note that the density does not go to zero smoothly; instead it extends up to $\pm x_0$, after which it
drops discontinuously to zero.  
The value of $x_m$ is determined by the density normalization, $N=\int \rho(x) dx dy$, 
$x_m = \sqrt{\frac{2 \pi}{\alpha L_y x_0}-\frac{x_0^2}{3}}$, and the extent of the
density distribution by the  force balance, $q \varphi'(x_0) =-\alpha x_0$.  
Using the Green function (\ref{green}), the force balance on a particle at $x=x_0$ requires
\begin{equation}
\frac{1}{4} \int_{-x_0}^{x_0} (x_m^2 - x^2) e^{-(x_0-x)}dx=  x_0,
\end{equation}
which reduces to, 
$-2+2 x_0-x_0^2+x_m^2+e^{-2 x_0} \left(2+2 x_0+x_0^2-x_m^2\right)=4 x_0$. 
\begin{figure}[h]
\begin{center}
\includegraphics[width=8cm]{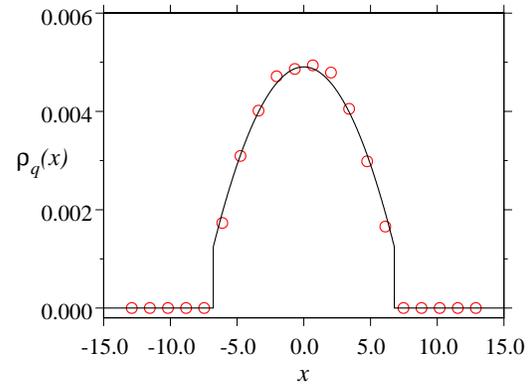}
\end{center}
\caption{The vertex strength density distribution $\rho_q(x) \equiv q^2 \rho(x)$ for $\alpha=10^{-3}$ and $L_y=20$.  
Circles are the result of our MD simulation with $N=1000$ particles, while the
solid curve is the prediction of the present theory. There are no adjustable parameters.}
\label{fig1}
\end{figure}

This constitutes the exact solution for the particle distribution at $T=0$.  
In Fig. 1 we compare it with the MD simulation.  As expected, an excellent agreement
is found between the theory and the simulation, without any fitting parameters.  
We conclude, therefore,  that the density distribution 
of particles in contact with a reservoir at $T=0$ has nothing to do with the Tsallis statistics, and everything
to do with Newton's Second Law. Of course for finite temperature, the density distribution will be
described by the usual Maxwell-Boltzmann statistical mechanics. 

This work was partially supported by the CNPq, INCT-FCx, and by the US-AFOSR under the grant FA9550-09-1-0283.

\noindent
Yan Levin and Renato Pakter\\
Instituto de Física, UFRGS\\ CP 15051, 91501-970, Porto Alegre, RS,
Brazil

\end{document}